\documentclass{appolb}
\usepackage{epsfig}

\begin{document}
\title{PARTICLE-HOLE MULTIPLETS NEAR CLOSED SHELLS%
\thanks{Presented at the XXXVII Zakopane School of Physics ``Trends in Nuclear Physics",
Zakopane, Poland, September 3-10, 2002 } }%


\author{A. Covello, L. Coraggio, A. Gargano, and N. Itaco
\address{Dipartimento di Scienze Fisiche, Universit\`a di Napoli Federico II \\
and Istituto Nazionale di Fisica Nucleare \\ Complesso Universitario di Monte S. Angelo \\ Via Cintia, I-80126 Napoli, Italy} }

\maketitle

\begin{abstract}
We report here on a shell-model study of nuclei close to doubly magic $^{132}$Sn and $^{100}$Sn focusing attention on particle-hole multiplets. In our study we make use of realistic effective interactions derived from the CD-Bonn nucleon-nucleon potential. We present results for the four nuclei $^{132}$Sb, $^{130}$Sb, $^{102}$In and $^{98}$Ag. Comparison shows that the calculated results are in very good agreement with the experimental data available for these nuclei far from stability. This supports confidence in the predictions of our calculations which may stimulate, and be helpful to, future experiments.

\end{abstract}
  
\section{Introduction}

The study of nuclei around double shell closures plays a key role in testing the predictive power of the shell model. In this context, of special interest are 
nuclei in the close vicinity to $^{208}$Pb, $^{132}$Sn and $^{100}$Sn, which may be assumed to be good closed cores. Experimentally, a large body of data is available in the $^{208}$Pb region, whereas it is a very hard task to obtain information on $^{132}$Sn and $^{100}$Sn neighbors. In recent years, however, substantial progress has been made to access the regions of shell closures off stability, which has paved the way to spectroscopic studies of nuclei in the neighborhood of both $^{132}$Sn and $^{100}$Sn. These data offer the opportunity for testing the basic ingredients of shell-model calculations, especially the matrix elements of the effective interaction, well away from the valley of stability. This may support confidence in the predictions of calculations, which may, in turn, stimulate efforts to gain more experimental information on these nuclei.

During the last few years, we have studied [1,2] several nuclei around  $^{132}$Sn and $^{100}$Sn within the framework of the shell model employing realistic effective interactions derived from modern nucleon-nucleon ($NN$) potentials. In these studies we have been mainly concerned with nuclei having few identical particles or holes.

The main aim of this paper is to report on some results of our current work in these regions, which we have obtained starting from the CD-Bonn free $NN$ potential \cite{machl01}. In particular, we shall consider the four odd-odd nuclei
$^{132}$Sb, $^{130}$Sb, $^{102}$In and $^{98}$Ag. The motivation for this choice lies in the fact that these nuclei are a source of information on the effective neutron-proton interaction in the $^{132}$Sn and $^{100}$Sn regions. More precisely, we focus attention on proton-neutron hole and neutron-proton hole multiplets in the former and latter region, respectively. 

To place this work in its proper perspectives, a brief historical digression is in order. More than thirty years ago the study of particle-hole multiplets in the $^{208}$Pb region was the subject of great experimental and theoretical 
interest [4-7]. In particular, the spectrum of $^{208}$Bi was extensively studied through pick-up and stripping reactions and several proton-neutron hole multiplets were identified [4,5]. It turned out that a main feature of all multiplets with more than two members is that the states with minimum and maximum $J$ have the highest excitation energy, while the state with next to the highest $J$ is the lowest one. On the theoretical side, a very good agreement with experiment was obtained in the shell-model study of Ref. \cite{kuo68}, where  particle-hole matrix elements deduced from the Hamada-Johnston $NN$ potential \cite{hama62} were employed. 

Despite these early achievements, little work along these lines has been done ever since. By considering the new data which are becoming available in the $^{132}$Sn and $^{100}$Sn regions and the prospect of spectroscopic studies of unstable nuclei opened up by the development of radioactive ion beams, we feel that the time is ripe to revive theoretical interest in  this subject and perform shell-model calculations making use of a modern $NN$ potential and improved many-body methods for deriving the effective interaction.

In Sec. 2 we give an outline of the theoretical framework in which our realistic shell-model calculations have been performed. In Sec. 3 we present several calculated particle-hole multiplets and show that all the available experimental data are well reproduced by our calculations. Sec. 4 presents some conclusions of our study.	

\section{Theoretical framework}

In our calculations for the Sb isotopes we assume that $^{132}$Sn is a closed core and let the valence proton and neutron holes occupy the five single-particle levels $0g_{7/2}$, $1d_{5/2}$, $1d_{3/2}$, $2s_{1/2}$, and $0h_{11/2}$ of the 50-82 shell.
Similarly, for $^{102}$In and $^{98}$Ag we assume these orbits are occupied by the valence neutrons outside the $^{100}$Sn core while for the proton holes the model space includes the four orbits $0g_{9/2}$, $1p_{1/2}$, $1p_{3/2}$, and $0f_{5/2}$ of the 28-50 shell.
For the Sb isotopes the single-proton and single neutron-hole energies have been taken from the experimental spectra [9-11] of $^{133}$Sb  and $^{131}$Sn, respectively. The only exception is the proton $\epsilon_{s_{1/2}}$ which was taken from Ref. \cite{andr97}, since the corresponding single-particle level is still missing in $^{133}$Sb. For $^{102}$In and $^{98}$Ag the single-particle and single-hole energies cannot be taken from experiment, since no spectroscopic data are yet available for $^{101}$Sn and $^{99}$In. Therefore, we have taken them from Ref. \cite{andr96} and Ref. \cite{cor00}, respectively, where they were determined by an analysis of the low-energy spectra of the Sn isotopes with $A \leq 111$ and of the $N=50$ isotones with $A\geq 89$. 

As already mentioned in  the Introduction, in our shell-model calculations we have made use of a realistic effective interaction derived from the CD-Bonn free nucleon-nucleon potential \cite{machl01}. This high-quality $NN$ potential, which is based upon meson exchange, fits very accurately ($\chi^2$/datum $\approx 1$) the world $NN$ data below 350 MeV available in the year 2000.

The shell-model effective interaction $V_{\rm eff}$ is defined, as usual, in the following way. In principle, one should solve a nuclear many-body Schr\"odinger equation of the form 
\begin{equation}
H\Psi_i=E_i\Psi_i, 
\end{equation}
with $H=T+V_{NN}$, where $T$ denotes the kinetic energy. This full-space many-body problem is reduced to a smaller model-space problem of the form
\begin{equation}
PH_{\rm eff}P \Psi_i= P(H_{0}+V_{\rm eff})P \Psi_i=E_iP \Psi_i.
\end{equation}
Here $H_0=T+U$ is the unperturbed Hamiltonian, $U$ being an auxiliary potential introduced to define a convenient single-particle basis, and $P$ denotes the projection operator onto the chosen model space.

A main difficulty encountered in the derivation of $V_{\rm eff}$ from a modern $NN$ potential, such as CD-Bonn, is the existence of a strong repulsive core which prevents its direct use in nuclear structure calculations. This difficulty is usually overcome by resorting to the well known Brueckner $G$-matrix method. Here, we have made use of a new approach \cite{bogn02} which has proved to be an advantageous alternative [15,16] to the use of the above method. The basic idea underlying this approach is to construct a low-momentum $NN$ potential, $V_{low-k}$, that preserves the physics of the original potential $V_{NN}$ up to a certain cut-off momentum $\Lambda$. In particular, the scattering phase shifts and deuteron binding energy calculated by $V_{NN}$ are reproduced by $V_{low-k}$. The latter is a smooth potential that can be used directly as input for the calculation of shell-model effective interactions. A detailed description of our derivation of $V_{low-k}$ can be found in Ref. \cite{bogn02}, where a criterion for the choice of the cut-off parameter $\Lambda$ is also given. We have used here the value $\Lambda=2.1$ fm$^{-1}$. 

Once the $V_{low-k}$ is obtained, the calculation of the matrix elements of the particle-particle and hole-hole interaction  is carried out within the framework of a folded-diagram method, as described, for instance, in Refs. [14,17].
It should be pointed out that the effective interaction in the proton-neutron channel has been explicitly derived in the particle-hole formalism. 
A description of the derivation of the particle-hole effective interaction is given in Ref. \cite{cor132}.

\section{Results an comparison with experiment}

We report here some selected results of our study of nuclei in the regions of shell closures off stability. The calculations have been performed using the OXBASH shell-model code \cite{OXBASH}.

We start by considering the Sb isotopes. The most appropriate system to study the proton-neutron interaction is $^{132}$Sb which, having one proton valence particle and one neutron valence hole, plays in the $^{132}$Sn region the role played by $^{208}$Bi in the Pb region.  Experimental information on this nucleus is provided by the studies of Refs. [19-21].
Two $\beta$-decaying isomers with $J^{\pi}=4^+$ and $8^-$ are known, which originate from the $\pi g_{7/2}$ $\nu d^{-1}_{3/2}$ and 
$\pi g_{7/2}$ $\nu h^{-1}_{11/2}$ configurations, respectively.
While the relative energy of these two states has not been determined, there are indications
that the $8^-$ state is located about 200 keV above the $4^+$ ground state \cite{ston89}.
  
As regards the positive-parity states, besides the other three members of the 
$\pi g_{7/2}$ $\nu d^{-1}_{3/2}$ multiplet, some states originating from the
$\pi g_{7/2}$ $\nu s^{-1}_{1/2}$, $\pi d_{5/2}$ $\nu d^{-1}_{3/2}$, and  $\pi g_{7/2}$ $\nu d^{-1}_{5/2}$ configurations have been identified. For the negative-parity states, very little experimental information is presently available. In a very recent study \cite{bhat01}, several new transitions feeding the $8^-$ isomer have been revealed. In particular, a $9^-$ state at 1.025 MeV has been identified which may be attributed to the $\pi g_{7/2}$ $\nu h^{-1}_{11/2}$  configuration. This makes very interesting the study of $^{130}$Sb, for which 
five low-lying states with $J^\pi=5^-,6^-,7^-,8^-$ and $9^-$ have been identified  [22,23] and interpreted as members of the same  multiplet. 

\begin{figure}[htb]
\begin{center}
\epsfxsize=10cm   
\epsfbox{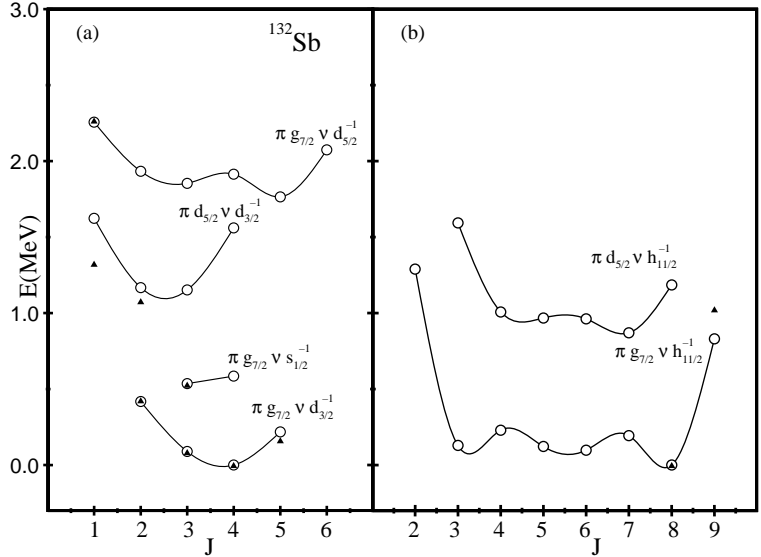}
\caption{Proton particle-neutron hole multiplets in $^{132}$Sb. The theoretical results are represented by open circles while the experimental data by solid triangles. The lines are drawn to connect the points.} 
\end{center}
\end{figure}

Several calculated multiplets for $^{132}$Sb are reported in Fig. 1 and compared with the existing experimental data.
Note that in Fig. 1(a) all energies are relative to the $4^+$ state while in Fig. 1(b) they are relative to the $8^-$ state (our calculations predict that the $8^-$ state lies 226 keV above the  $4^+$ ground state).

As regards the comparison between theory and experiment, we see that the calculated energies are in good agreement with
the observed ones. In fact, the discrepancies are all in the order of few tens of keV, the only exceptions being the
$1^+$  and $9^-$ states of the $\pi d_{5/2}$ $\nu d^{-1}_{3/2}$ and $\pi g_{7/2}$ $\nu h^{-1}_{11/2}$ multiplets, which come about 300 keV above and 200 keV below their experimental counterparts, respectively.

The calculated $\pi g_{7/2}$ $\nu h^{-1}_{11/2}$ multiplet for $^{130}$Sb is reported and compared  with the experimental data in Fig. 2.
We see that the agreement between experiment and theory is remarkably good, thus providing further support to our predictions for $^{132}$Sb.  It should be noted that both the experimental and calculated energies are relative to the $8^-$ state, which has been experimentally observed to be the ground state. Our calculations
predict \cite{cor132} for the ground state $J^\pi=4^+$, the $8^-$ state lying at 92 keV excitation energy.
As regards the wave functions of the states reported in Fig. 2, we find that they are indeed dominated by the $ph$ configuration  $\pi g_{7/2}$ $\nu h^{-1}_{11/2}$ while the remaining two 
neutron holes give rise to a zero-coupled pair occupying the $d_{3/2}$, $h_{11/2}$, and $s_{1/2}$ levels. 

\begin{figure}[htb]
\begin{center}
\epsfbox{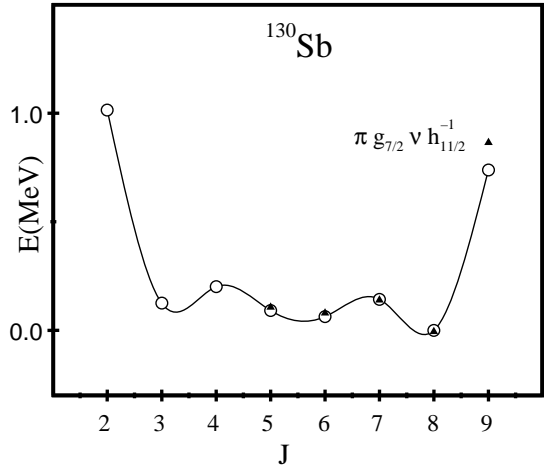}
\caption{Same as Fig.1, but for the $\pi g_{7/2}$ $\nu h^{-1}_{11/2}$ multiplet in $^{130}$Sb.} 
\end{center}
\end{figure}

It is evident from Figs. 1 and 2 that a main feature of the calculated multiplets is that the states with minimum and maximum $J$ have the highest excitation energy
and are well separated from the other states, for which the splitting is relatively small. This pattern is in agreement with the experimental one for the $\pi g_{7/2}$ $\nu d^{-1}_{3/2}$ multiplet in $^{132}$Sb and the experimental data available for the other multiplets also go in the same direction. It should be pointed out that this behavior  is quite similar to that exhibited [6,7] by the multiplets in the heavier particle-hole nucleus $^{208}$Bi. Also, it is worth noting that in all of our calculated multiplets the state of spin ($j_\pi + j_\nu -1$) is the lowest, in agreement with the early predictions of the Brennan-Bernstein coupling rule \cite{bren60}.

We turn now to the $^{100}$Sn region. In this region the counterpart of $^{208}$Bi and  $^{132}$Sb is $^{100}$In, for which studies of excited states are at present out of reach. We have therefore considered the two neighboring odd-odd isotopes $^{102}$In and $^{98}$Ag, focusing attention on the 
$\pi g_{9/2}^{-1}\nu d_{5/2}$ multiplet for which some experimental information is available.
In Figs. 3 and 4 we report the results of our calculations for $^{102}$In and $^{98}$Ag, respectively, and compare them with the experimental data \cite{nndc}. We see that the agreement between experiment and theory is of the same quality as that obtained in the $^{132}$Sn region, the largest discrepancy being 130 keV for the $5^+$ state in $^{98}$Ag.
The pattern of the calculated multiplets turns out to be similar to that of the multiplets in the Sb isotopes, but with the states with minimum and maximum $J$ less separate from the other ones. Actually, for $^{102}$In we find that the energy of the $7^+$ state is about the same as that of the $3^+$ state. In this connection, it should be mentioned that for the former state the percentage of configurations other than those having a $g_{9/2}$ proton hole and a $d_{5/2}$ neutron is about 50\%.  
A detailed discussion of the structure of the calculated states will be included in  a forthcoming publication.

\begin{figure}[ht]
\begin{center}
\epsfbox{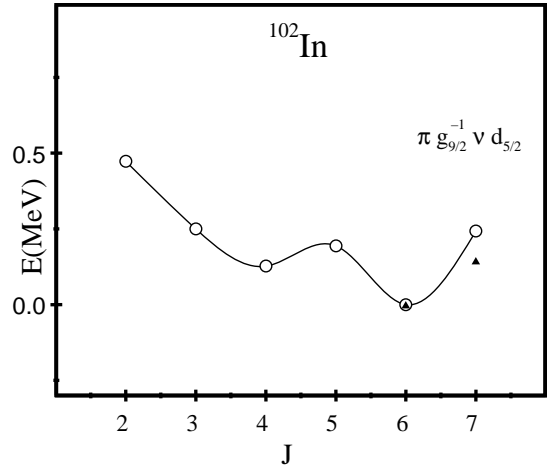}
\caption{Same as Fig. 1, but for  the $\pi g_{9/2}^{-1}$ $\nu d_{5/2}$ multiplet in $^{102}$In.} 
\end{center}
\end{figure}

\begin{figure}[ht]
\begin{center}
\epsfbox{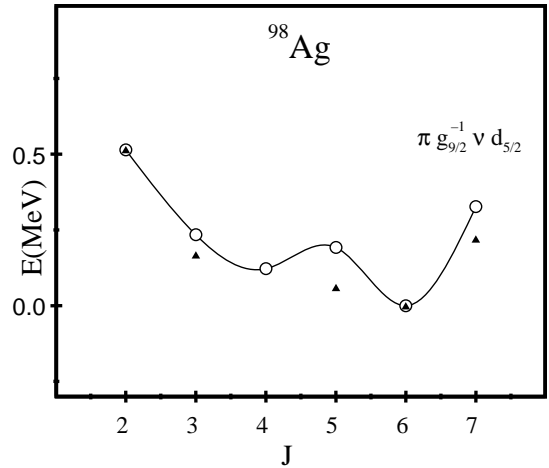}
\caption{Same as Fig. 1, but for  the  $\pi g_{9/2}^{-1}$ $\nu d_{5/2}$ multiplet in $^{98}$Ag.} 
\end{center}
\end{figure}

\section{Summary}

We have presented here some results of a shell-model study of nuclei in the close vicinity to doubly magic $^{132}$Sn and $^{100}$Sn focusing attention on the particle-hole multiplets,
which are a direct source of information on the neutron-proton effective interaction.
In our calculations we have employed a realistic effective interaction derived from the CD-Bonn $NN$ potential. This has been done using a new 
approach \cite{bogn02} to shell-model effective interactions instead of the  usual Brueckner $G$-matrix method. 

We have shown that our results are in very good agreement with the available experimental data. It should be stressed that in our calculations no use has been made of any adjustable parameter.
In fact, for the single-neutron and single proton-hole energies in the $^{100}$Sn region, which are not available from experiment, we have adopted the values determined in our earlier studies [13,14] on Sn isotopes and $N=50$ isotones.

In summary, we may conclude that the results of the present study of particle-hole multiplets in the $^{132}$Sn and $^{100}$Sn regions are consistent with those obtained in our previous studies of nuclei with identical valence particles and holes in the same regions \cite{cov01}. This  confirms that realistic shell-model calculations are able to describe with quantitative accuracy the spectroscopic properties of nuclei in regions of shell closures off stability,
giving confidence in their predictive power. On the experimental side, it is of key importance to gain more information on nuclei in these  regions. This is certainly a main challenge  of nuclear structure experiments with radioactive ion beams. 

\vskip 0.5cm 

This work was supported in part by the Italian Ministero dell'Istruzione, dell'Universit\`a e della Ricerca (MIUR).

\end{document}